\documentstyle[preprint,prd,aps,psfig]{revtex}\tighten
\newcommand{\beq}{\begin{equation}}
\newcommand{\beqa}{\begin{eqnarray}}
\newcommand{\eeq}{\end{equation}}
\newcommand{\eeqa}{\end{eqnarray}}
\newcommand{\p}{\phi}

\begin{document}
\draft
\preprint{\tighten\vbox{\hbox{KUNS-1791}\hbox{astro-ph/0206298}}}

\title{Tracking K-essence}
\author{Takeshi Chiba}

\address{Department of Physics, Kyoto University, 
Kyoto 606-8502, Japan}

\date{\today}

\maketitle

\bigskip

\begin{abstract}
 We derive a condition for converging a common evolutionary
 track for k-essence (a scalar field dark energy with non-canonical 
 kinetic terms). For the Lagrangian density  $V(\phi)W(X)$ with 
 $X=\dot\p^2/2$, we find tracker solutions with $w_{\p} < w_B$ 
 exist if $\Gamma\equiv V''V/(V')^2 > 3/2$. Here $w_{\p}(w_B)$ is the
 equation-of-state of the scalar field (background radiation/matter). 
 Our condition may be useful for examining the existence of the
 attractor-like behavior in cosmology with k-essence (for example,
 rolling tachyon). 
\end{abstract}


\section{introduction}

Scalar fields play important role in cosmology. For example, 
inflation in the early universe may be caused by inflatons, 
or an ultra-light scalar field with mass $\sim 10^{-33}$ eV (or
quintessence) may cause the universe to accelerate recently. 

In the case of inflation models, the conditions for existence of 
inflationary solutions are conveniently described in terms of the slow-roll
parameters without solving the equation of motion directly:  
$(V'/V)^2/2\kappa^2 <1$ and $V''/\kappa^2V <1$ with $\kappa^2=8\pi G$ 
\cite{rmp}.

Similarly, some quintessence models admit so called tracker fields which 
have attractor-like solutions in the sense that a very wide range of
initial conditions rapidly converge to a common cosmic evolutionary track
\cite{swz}. 
It may be particularly useful to express the condition for the existence
of tracking solutions in terms of a simple condition of $V(\phi)$
without having to solve the equation of motion directly. For
quintessence, the condition for the existence of tracker solutions with
$w_{\p}< w_B$ is $\Gamma\equiv V''V/(V')^2>1$ \cite{swz}. Here
$w_{\p}(w_B)$ is the equation-of-state of quintessence (background
radiation or matter). 

Usually the quintessence field (or inflaton) is modeled by a scalar 
field with a canonical kinetic term and a potential term. However, as
shown in \cite{adm,coy}, a scalar field with solely kinetic terms can 
(even without potential terms), albeit they are non-canonical, mimic
such a (canonical) quintessence/inflaton field.\footnote{We named such 
quintessence as ``kinetic'' quintessence in \cite{coy}. 
However, later the name ``k-essence'' was coined by the authors 
of \cite{ams} and got popularity. So in this paper we reluctantly use
the term k-essence for such kinetically driven quintessence.}
However, the condition for the existence of tracker solutions 
for k-essence is not known to date. 

In view of recent interest in scalar field 
cosmology with non-canonical kinetic terms \cite{sen,tachyon,tachyon2}, 
in this paper we shall express the condition for tracker solutions in 
terms of a function of the Lagrangian density. 

\section{Tracking K-essence}

\subsection{Basics of K-essence}

The action of  K-essence minimally coupled with gravity is
\beq
S = \int d^4x\sqrt{-g}\left({1\over 2\kappa^2}R+p(\p,X)\right),
\eeq
where $\kappa^2=8\pi G$ and $X=-\nabla^{\mu}\p\nabla_{\mu}\p/2$. 
The pressure of the scalar
field $\phi$, $p_{\p}$, is given by $p(\p,X)$ itself and the energy
density $\rho_{\p}$ is given by $\rho_{\p}=2X\partial p/\partial X-p$
\cite{adm,coy}.

The field equations in flat Friedmann-Robertson-Walker spacetime are
\beqa
&&H^2 := \left({\dot a\over a}\right)^2
= \frac{\kappa^2}{3}(\rho_B+\rho_{\p})
= \frac{\kappa^2}{3}\left(\rho_B+2X{\partial p\over\partial
    X}-p\right), \label{eq:hubble} \\
&&{\ddot{a}\over a} = -{\kappa^2\over
  6}(\rho_B+3p_B+\rho_{\p}+3p_{\p})=:-{\kappa^2\over 6}
\left((1+3w_B)\rho_B+(1+3w_{\p})\rho_{\p}\right),
\label{eq:a} \\
&&\ddot{\p}\left({\partial p\over \partial X}+
  \dot\p^{2}{\partial^{2}p\over \partial X^2}\right)+3H{\partial
  p\over \partial X}\dot\p+{\partial^{2}p\over \partial X\partial
  \p}\dot\p^{2}-{\partial p\over \partial\p} = 0,
\label{eq:q_eom1}
\eeqa
where $\rho_B$ and $p_B$ are the energy density and the pressure of
the background matter and/or radiation, respectively. 

In this paper, as a first step toward more general case, 
we shall derive a tracker condition for a k-essence with 
the following factorized form of $p(\p,X)$:
\beq
p(\p,X)=V(\p)W(X).
\label{lagrangian}
\eeq
This form of the Lagrangian is suggested for that of tachyon by using 
the boundary string field theory \cite{kmm,st}. Moreover, any
Lagrangian containing only $\dot\phi^2$ and $\dot\phi^4$ terms can be recast
in the factorized form after field redefinition.

The equation of motion of the scalar field is then written as
\beq
\ddot\p\left(W_X+2XW_{XX}\right)+3HW_X\dot\p+\left(2XW_X-W\right){V'\over
  V}=0,
\eeq
where $W_X=dW/dX$.

\subsection{Tracker Equation}

We can express the equation of motion of $\p$ in alternative form which
may be useful for the following analysis:
\beqa
{V'\over V^{3/2}}&=&\mp {\kappa \over 2}
\sqrt{{(1+w_{\p})W_X\over 3\Omega_{\p}} }
\left(6+   A y'\right),\label{eom}\\
A &=& {(XW_X-W)(2XW_{XX}+W_X)\over XW_X^2-WW_X-XWW_{XX}}={1-w_{\p}\over
    c_s^2-w_{\p}},
\eeqa
where $y=(1+w_{\p})/(1-w_{\p})$ and 
$ y' =d\ln y/d\ln a$, and minus(plus) sign corresponds to $\dot\p
>0(<0)$, respectively. $c_s^2$ is the speed of sound of k-essence defined
by \cite{gm}
\beq
c_s^2={p_X\over p_X+2Xp_{XX}}.
\eeq
Note that for quintessence with a canonical kinetic term, $c_s^2=1$. 
For a tracker solution ($w_{\p}\simeq $ const.),
we obtain a relation:
\beq
{1\over \sqrt{\Omega_{\p}}}=\mp {1\over \kappa\sqrt{3(1+w_{\p})W_X}}
{V'\over
      V^{3/2}},\label{trac:condition}
\eeq
which might be called the k-essential counterpart of the tracker
condition \cite{swz}. 

Similar to \cite{swz}, we define a dimensionless function $\Gamma$ 
by  $\Gamma =VV''/V'^2$. After taking 
the time derivative of Eq.(\ref{eom}), we obtain
\beqa
&&\Gamma-{3\over 2}=-{1\over (1+w_{\p})(6+A y')}\left[
3(w_{\p}-w_B)(1-\Omega_{\p}) +{(1-w_{\p})^2\over 
2(c_s^2-w_{\p})} y'\right. \nonumber\\
&&+\left.{2(1-w_{\p})(c_s^2-w_{\p}) y'' + 2\left(\dot w_{\p}(1-c_s^2)-
(c_s^2)^{\cdot}(1-w_{\p})\right) y'/H
\over (6+A y')(c_s^2-w_{\p})^2}
\right],\label{tracker}
\eeqa
where $ y''=d^2\ln y/d\ln a^2$. Eq.(\ref{tracker}) might be called 
the k-essential counterpart of the tracker equation.
Therefore for the tracker solution (assuming $\Gamma \simeq$ const. 
and $\Omega_{\p}\ll 1$\footnote{This assumption is implicit in \cite{swz}.}) 
we can write $w_{\p}$ in terms of $\Gamma$:
\beq
w_{\p}={w_B-2(\Gamma -3/2)\over 2(\Gamma -3/2)+1}\simeq {\rm const}.\label{eos}
\eeq

\subsection{Tracking Condition}

\paragraph*{Convergence toward the tracker solution.}
We now examine the stability of the tracker solution. Consider a
solution which is perturbed from the tracker solution with $w_0$
(Eq.(\ref{eos})) by an
amount $\delta$, then the tracker equation Eq.(\ref{tracker}) is
expanded to lowest order in $\delta$ to obtain
\beq
2\delta'' +3(1+w_B-2w_0)\delta' +9(c_s^2-w_0)(1+w_B)\delta=0,
\eeq
where the prime means $d/d\ln a$. The solution of this equation is 
\beqa
&&\delta \propto a^{\gamma}\\
&&\gamma=-{3\over 4}(1+w_B-2w_0)\pm {3\over
  4}\sqrt{(1+w_B-2w_0)^2-8(c_s^2-w_0)(1+w_B)}.
\eeqa
In order for the real part of $\gamma$ to be negative so that $\delta$
decays exponentially and the solution approaches the tracker solution, 
it is required that
\beq
w_0< {1+w_B\over 2}~~~~~{\rm and }~~~~w_0< c_s^2,
\label{condition}
\eeq
where $c_s^2\geq 0$ is assumed for stability against perturbation. 
Note that for the canonical quintessence ($c_s^2=1$) the second
requirement in Eq.(\ref{condition}) is automatically satisfied. 

{}From Eq.(\ref{eos}), the above requirements are written in terms of 
$\Gamma$:
\beqa
&& \Gamma> {3\over 2}-{c_s^2-w_B\over 2(c_s^2+1)}~~~{\rm and }~~~ 
\Gamma> {3\over 2}-{1-w_B\over 6+2w_B}\\
&&{\rm or}~~~ \Gamma < 1.
\eeqa

\paragraph*{Tracking behavior with $w_{\p} < w_B$.}
In this case, $\Omega_{\p}$ increases with increase of time. 
Then according to Eq.(\ref{trac:condition}), $|V'/V^{3/2}\sqrt{W_X}|$ 
decreases for a tracker solution. On the other hand, using the equation 
of motion, we find that
\beq
\left({V'\over V^{3/2}\sqrt{W_X}}\right)^{.}={V'\dot\p\over
  V^{5/2}\sqrt{W_X}}\left(\Gamma -{3\over 2}\right).
\eeq
Hence, $|V'/V^{3/2}\sqrt{W_X}|$ decreases if $\Gamma > 3/2$. 
Therefore, tracking behavior with $w_{\p} < w_B$ occurs if $\Gamma >
3/2$ and nearly constant. 
{}From the stability analysis, this tracker solution is stable if
$\Gamma > \max(3/2,3/2-(c_s^2-w_B)/2(c_s^2+1))$. In particular for
$w_B=0$, the condition is simply $\Gamma > 3/2$. 

\paragraph*{Tracking behavior with $w_{\p} > w_B$.}
This is possible for $1<\Gamma < 3/2$ and nearly constant. 
However, from the stability analysis, we find that 
$w_{\p}< \min((1+w_B)/2,c_s^2)$ is additionally required, which demands 
$\Gamma-3/2 > \max(-(c_s^2-w_B)/2(c_s^2+1),-(1-w_B)/(6+2w_B))$. 

\paragraph*{Tracking behavior with $w_{\p}< -1$.} 
This is possible for $\Gamma < 1$ and nearly constant. From the
stability analysis, we find that this solution is stable.

\subsection{Examples}

As an example, we consider the universe consisting of matter/radiation
and a k-essential scalar field with a power-law model, 
$V\propto \p^{-\alpha}$, studied in \cite{coy}. Since 
$\Gamma=(\alpha +1)/\alpha$, if $0<\alpha < 2$,
then tracking behavior with $w_{\p}<w_B$ occurs, while  if $\alpha < 0$, 
tracking behavior with $w_{\p}<-1$ occurs. From Eq.(\ref{eos}), 
in this case we have for the tracker solution
\beq
w_{\p}=(1+w_B){\alpha\over 2}-1,
\eeq
which indeed coincides with the equation-of-state of the attractor 
solution studied in \cite{coy} and indeed the solution is stable \cite{coy}. 

Another example is the universe consisting of matter/radiation and 
the rolling tachyon with $W(X)=-\sqrt{1-2X}$. The exact classical 
potential of it has been computed in \cite{kmm} and is given
by
\beq
V(T)=V_0\left(1+T/T_0\right)\exp(-T/T_0),
\eeq
where $T$ is the tachyon field and $V_0$ is the tension of some
unstable bosonic D-brane and $T_0$ is a constant of the order of the 
string scale. In this case we have
\beq
\Gamma=1- (T/T_0)^{-2}
\eeq
and $\Gamma$ becomes nearly constant if  $T/T_0 \gg 1$. 
In the limit of large $T/T_0$, we have $\Gamma = 1$ and 
 obtain $w_{\p}=-[w_B(1-\Omega_{\p})+1]/\Omega_{\p}$ 
if we do not neglect $\Omega_{\p}$ in Eq.(\ref{eos}). 
Since the rolling tachyon respects the weak energy condition, $w_{\p} \geq
-1$, this implies $\Omega_{\p} \geq 1$ during the radiation dominated 
epoch (the same result holds for the matter dominated epoch), 
which would be incompatible with the success of the Big-Bang
Nucleosynthesis. 
Thus the above potential does not admit viable tracker solutions, 
which implies the need of fine-tuning to account for $\Omega_{\p}\sim 1$ 
today in agreement with the recent analysis \cite{tachyon2}. 

\section{Summary}

We have derived a (sufficient) condition for the existence of tracker
solutions for the system of matter/radiation and a scalar field 
with Lagrangian density of the form Eq.(\ref{lagrangian}). 
Our results are summarized as follows:

\begin{itemize}

\item Tracking behavior with $w_{\p} < w_B$ occurs if 
$\Gamma > \max(3/2,3/2+(w_B-c_s^2)/2(c_s^2+1))$ and nearly constant. 
In particular for $w_B=0$, the condition is simply $\Gamma > 3/2$ and
nearly constant. 

\item Tracking behabior with $w_{\p}>w_B$ occurs if 
$3/2>\Gamma > 3/2+\max(-(c_s^2-w_B)/2(c_s^2+1),-(1-w_B)/(6+2w_B))$ and
nearly constant. 

\item Tracking behavior with  $w_{\p}<-1$ occurs if $\Gamma <1$ and
  nearly constant.

\end{itemize}

Interestingly, thanks to the factorizable ansatz Eq.(\ref{lagrangian}), 
the conditions are similar to those of canonical quintessence. 

\acknowledgments

We would like to thank Masahide Yamaguchi for useful comments. 
This work was supported in part by a Grant-in-Aid for Scientific 
Research (No.13740154) from the Japan Society for the Promotion of
Science and by a Grant-in-Aid for Scientific Research on Priority Areas 
(No.14047212) from the Ministry of Education, Science, Sports and 
Culture, Japan.


\end{document}